\documentclass[num-refs]{nbdt-article}

\usepackage{siunitx}

\papertype{}
\title{Golden rhythms as a theoretical framework for cross-frequency organization}

\author[1,2\authfn{1}]{Mark A. Kramer}

\affil[1]{Department of Mathematics and Statistics, Boston University}
\affil[2]{Center for Systems Neuroscience, Boston University}

\corraddress{Mark A. Kramer, Department of Mathematics and Statistics, Boston University, Boston, MA, 02115, USA}
\corremail{mak@bu.edu}

\fundinginfo{NIH, NIBIB, Grant/Award Number: R01EB026938; NSF, Grant/Award Number: 1451384}

\runningauthor{Kramer, MA}

\begin{document}

\maketitle

\begin{abstract}
While brain rhythms appear fundamental to brain function, why brain rhythms consistently organize into the small set of discrete frequency bands observed remains unknown. Here we propose that rhythms separated by factors of the golden ratio ($\phi=(1+ \sqrt{5})/2$) optimally support segregation and cross-frequency integration of information transmission in the brain. Organized by the golden ratio, pairs of transient rhythms support multiplexing by reducing interference between separate communication channels, and triplets of transient rhythms support integration of signals to establish a hierarchy of cross-frequency interactions. We illustrate this framework in simulation and apply this framework to propose four hypotheses.

% Please include a maximum of seven keywords
\end{abstract}

\vfil
\centering
{\Large In memory of Prof. Miles A. Whittington}

\clearpage

\justifying

\section{Introduction}
The brain is organized into a hierarchy of functionally specialized regions, which selectively coordinate during behavior \cite{hipp_oscillatory_2011,vezoli_brain_2021,engel_dynamic_2001,salinas_correlated_2001} and rest \cite{fox_spontaneous_2007,brookes_investigating_2011,de_pasquale_cortical_2012}. Effective function relies on dynamic coordination between brain regions, in response to a changing environment, on an essentially fixed and limited anatomical substrate \cite{kopell_beyond_2014,quiroga_closing_2020,tononi_information_2004,kohn_principles_2020}. Through these anatomical connections multiplexing occurs: multiple signals that combine for transmission through a single communication channel must then be differentiated at a downstream target location \cite{akam_oscillatory_2014,becker_resolving_2020}. How information – communicated via coordinated transmission of spiking activity \cite{hahn_portraits_2019} – dynamically routes through the brain’s complex, distributed, hierarchical network remains unknown \cite{palmigiano_flexible_2017}. 

\noindent Brain rhythms – approximately periodic fluctuations in neural population activity – have been proposed to control the flow of information within the brain network \cite{akam_oscillatory_2014, fries_mechanism_2005,gonzalez_communication_2020,canolty_functional_2010,bonnefond_communication_2017,buzsaki_brain_2012} and proposed as the core of cognition \cite{palva_discovering_2012,siegel_spectral_2012,siebenhuhner_genuine_2020,williams_modules_2021}. Through periodic modulations in neuronal excitability, rhythms may support flexible and selective communication, allowing exchange of information through coordination of phase at rhythms of the same frequency (e.g., coherence \cite{fries_mechanism_2005,bonnefond_communication_2017,bastos_communication_2015,fries_rhythms_2015,schroeder_low-frequency_2009}) and different frequencies (e.g., phase-amplitude coupling \cite{canolty_functional_2010,lisman_theta-gamma_2013,hyafil_neural_2015} or n:m phase locking \cite{palva_phase_2005,belluscio_cross-frequency_2012,tass_detection_1998}).  Recent evidence shows that neural oscillations appear as transient, isolated events \cite{lundqvist_gamma_2016,sherman_neural_2016}; how such transient oscillations route information through neural networks remains unclear \cite{van_ede_neural_2018}. 

\noindent Significant evidence supports the organization of brain rhythms into a small set of discrete frequency bands (e.g., theta [4-8 Hz], alpha [8-12 Hz], beta [12-30 Hz], gamma [30-80 Hz]) \cite{buzsaki_rhythms_2011,buzsaki_neuronal_2004}. Consistent frequency bands appear across mammalian species (mouse, rat, cat, macaque, and humans \cite{buzsaki_scaling_2013}) and in some cases the biological mechanisms that pace a rhythm are well-established (e.g., the decay time of inhibitory post-synaptic potentials sets the timescale for the gamma rhythm \cite{whittington_inhibition-based_2000}). Why brain rhythms organize into discrete bands, and whether these rhythms are fixed by the brain’s biology or organized to optimally support brain communication, remains unclear. For example, an alternative organization of the brain’s rhythms (e.g., into a larger set of different frequency bands) may better support communication but remain inaccessible given the biological mechanisms available to pace brain rhythms.

\noindent While much evidence supports the existence of brain rhythms and their importance to brain function, few theories explain their arrangement. Different factors have been proposed for the spacing between the center frequencies of neighboring bands: Euler’s number ($e\approx2.718$) \cite{penttonen_natural_2003}, the integer 2 \cite{klimesch_algorithm_2013}, or the golden ratio ($\phi\approx1.618$) \cite{roopun_temporal_2008}. Existing theory shows that irrational factors (e.g., $e$ and $\phi$) minimize interference between frequency bands, in support of separate rhythmic communication channels for multiplexing information in the brain \cite{izhikevich_weakly_1997,hoppensteadt_thalamo-cortical_1998,izhikevich_weakly_1999}. However, if separate rhythmic channels communicate different information, and the organization of brain rhythms prevents interference, how a target location coordinates information across these rhythms is unclear. For example, how in theory a neural population integrates top-down and bottom-up input communicated in separate rhythmic channels (lower [$<$40 Hz] and higher [$>$40 Hz] frequency ranges, respectively \cite{bastos_communication_2015,fries_rhythms_2015,michalareas_alpha-beta_2016,fontolan_contribution_2014,bastos_layer_2020}) remains unclear. We propose a solution to this problem: addition of a third rhythm. Motivated by an existing mathematical theory \cite{izhikevich_weakly_1997,hoppensteadt_thalamo-cortical_1998,izhikevich_weakly_1999}, we show that effective communication among three rhythms is optimal for rhythms arranged according to the golden ratio.

\noindent In what follows, we show that golden rhythms – rhythms organized by the golden ratio – are the optimal choice to integrate information among separate rhythmic communication channels. We propose that brain rhythms organize in the discrete frequency bands observed, with the specific spacing observed, to optimize segregation and integration of information transmission in the brain.

\section{Methods}

All simulations and analysis methods to reproduce the manuscript results and figures are available at \url{https://github.com/Mark-Kramer/Golden-Framework}.

\subsection{Damped harmonic oscillator model} \label{meth:dsho}
As a simple model of rhythmic neural population activity (e.g., observed in the local field potential (LFP) or magneto/electroencephalogram (M/EEG)) we implement a network of coupled damped harmonic oscillators \cite{moorman_golden_2007}. We choose the damped harmonic oscillator for three reasons. First, a harmonic oscillator (e.g., a spring) mimics the restorative mechanisms governing displacements about a stable equilibrium in neural dynamics (e.g., excitation followed by inhibition in the gamma rhythm \cite{whittington_inhibition-based_2000,fries_gamma_2007}, depolarization followed by hyperpolarization – and vice versa – in bursting rhythms \cite{izhikevich_synchronization_2001}). Second, brain rhythms are transient \cite{lundqvist_gamma_2016,sherman_neural_2016}. In the model, damping (e.g., friction) produces transient oscillations that decay to a stable equilibrium. Third, the damped harmonic oscillator driven by noise is equivalent to an autoregressive model of order two (AR(2), see Appendix \ref{Appendix1}). The AR(2) model simulates stochastic brain oscillations \cite{spyropoulos_spontaneous_2020}, consistent with the concept of a neural population with resonant frequency driven by random inputs.
\\

\noindent We simulate an 8-node network of damped, driven harmonic oscillators. We model the activity $x_k$ at node $k$ as,
\begin{equation}\label{eq:dsho}
 \ddot{x}_k + 2 \beta \dot{x}_k + \omega_k^2 x_k = \left( \bar{g}_C + \bar{g}_S\cos{\omega_S t} \right) \sum_{j \neq k} x_j \ ,
\end{equation}
where $\beta$ is the damping constant, and $\omega_k=2 \pi f_k$ is the natural frequency of node $k$. The activity $x_j$ summed from all other nodes ($j \neq k$) drives node $k$. We modulate this drive by a gain function with two terms: a constant gain $\bar{g}_C$ and a sinusoidal gain with amplitude $\bar{g}_S$ and frequency $\omega_S=2 \pi f_S$. To include noise in the dynamics, we represent the second order differential equation in Equation (\ref{eq:dsho}) as two first order differential equations for the position and velocity of the oscillator. We add to the position dynamics a noise term, normally distributed with mean zero and standard deviation equal to the average standard deviation of the evoked response at all oscillators simulated without noise, excluding the perturbed oscillator from the average. In this way, we add meaningful noise of the same magnitude to all oscillators. We numerically simulate the model with noise using the Euler-Maruyama method. To examine the impact of different noise levels, we multiply the noise term by factors $\{0, 0.5, 1.0, 1.5, 2.0\}$. For each noise level, we repeat the simulation 100 times with random noise instantiations.

\section{Results}
In what follows, we propose that brain rhythms organized according to the golden ratio produce triplets of rhythms that establish a hierarchy of cross-frequency coupling. We conclude with four hypotheses deduced from this framework and testable in experiments.

\subsection{Rhythms organized by the golden ratio support selective cross-frequency coupling}

\begin{figure}[h]
\centering
\includegraphics[width=6cm]{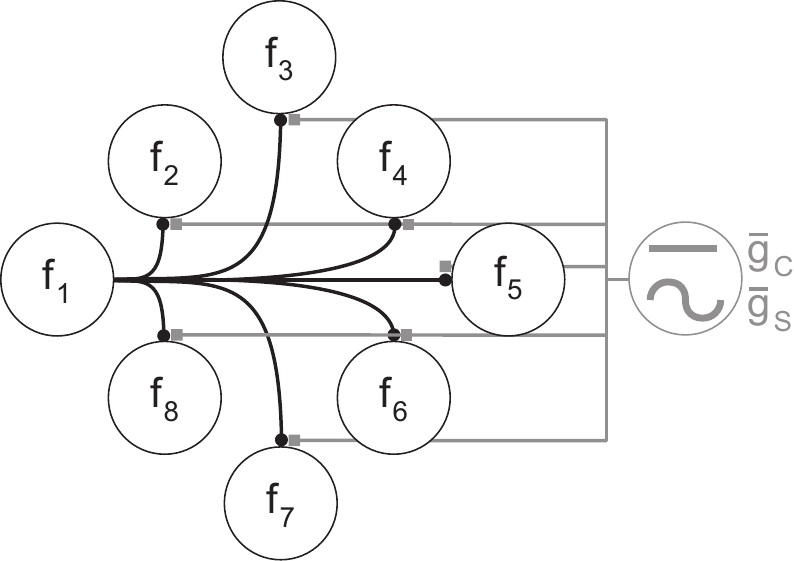}
\caption{\textbf{Illustration of the coupled oscillator network.} Oscillators with frequency $f_k$ receive input from all other oscillators. Input from one oscillator (frequency $f_1$) to all other oscillators $(f_2,f_3,\ldots,f_8)$ is shown (black curves); similar connectivities exist from all other oscillators (not shown). Constant ($\bar{g}_C$) and sinusoidal ($\bar{g}_S$) gain modulates each input (gray lines).} \label{fig:schematic}
\end{figure}

In the case of weakly-connected oscillatory populations, whether the populations interact or not depends on their frequency ratios \cite{izhikevich_weakly_1997,hoppensteadt_thalamo-cortical_1998,izhikevich_weakly_1999}; rational frequency ratios support interactions, while irrational frequency ratios do not. Motivated by this theory, we consider a network of interacting, rhythmic neural populations (Figure \ref{fig:schematic}). We model each population as a damped harmonic oscillator, with each oscillator assigned a natural frequency $f_k$. To couple the populations, we drive each oscillator with the summed activity of all other oscillators (i.e., the connectivity is all-to-all). We modulate this drive by a gain function ($g$) with constant ($\bar{g}_C$) and sinusoidal (amplitude $\bar{g}_S$, frequency $f_S$) terms: $g=\bar{g}_C+\bar{g}_S \cos{\left(2\pi f_S t \right)}$; see {\it Methods}. Analysis of this coupled oscillator system reveals resonance (i.e., a large amplitude response) at a target oscillator in two cases. To describe these cases, we denote the frequency of a target oscillator as $f_T$ and the frequency of a driver oscillator as $f_D$. A large amplitude (resonant) response occurs at the target oscillator in the following cases,
\\

constant gain modulation:
\vspace{-0.15in}
\begin{eqnarray}
    0 = f_T - f_D \label{eq:const}
\end{eqnarray}

sinusoidal gain modulation:
\vspace{-0.15in}
\begin{subequations}  \label{eq:sin}
\begin{eqnarray}
    f_S = f_T - f_D \label{eq:sin_a} \\
    f_S = f_D - f_T \label{eq:sin_b} \\
    f_S = f_D + f_T \label{eq:sin_c}
\end{eqnarray}
\end{subequations}

\noindent The first case (Equation \ref{eq:const}) corresponds to the standard result for a damped target oscillator driven by sinusoidal input; when the sinusoidal driver frequency $f_D$ matches the natural frequency of the target $f_T$, the response amplitude at the target is largest (e.g., see Chapter 5 of \cite{taylor_classical_2005}). The next three cases (Equation \ref{eq:sin}) correspond to a damped target oscillator driven by sinusoidal input modulated by sinusoidal gain. If the gain frequency $f_S$ equals the sum or difference of the target and driver frequencies, then the response amplitude at the target is largest (see Appendix \ref{Appendix2}). We note that the first case corresponds to within-frequency coupling (i.e., the driver and target have the same frequency) while the next three cases correspond to cross-frequency coupling (i.e., the driver and target have different frequencies). We also note that, in this model, we assume an oscillator responds to an input by exhibiting a large amplitude response; in this way, we consider the oscillation amplitude as encoding information, consistent with notion of information encoded in firing rate modulations \cite{akam_oscillations_2010}.

\noindent The results in Equations (\ref{eq:const}, \ref{eq:sin}) hold for any choice of driver, target, and gain frequencies without additional restrictions. We now apply an additional restriction, and consider the damped harmonic oscillator network with oscillator and gain frequencies $f_k$ satisfying,
\begin{eqnarray} \label{eq:fk}
    f_k = f_0 \, c^k \, ,
\end{eqnarray}
where $f_0 > 0$ determines the frequency at $k=0$. As discussed above, candidate values for $c$ deduced from {\it in vivo} observations include Euler’s number ($e\approx2.718$) \cite{penttonen_natural_2003}, the integer 2 \cite{klimesch_algorithm_2013}, or the golden ratio ($\phi\approx1.618$) \cite{roopun_temporal_2008}. Then, given the set of three neighboring frequencies $\{f_k,f_{k+1},f_{k+2}\}$, what choice of $c$ supports cross-frequency coupling in the network? To answer this, we choose $f_S = f_{k+2}$, $f_D = f_{k+1}$, and $f_T = f_k$ so that Equation (\ref{eq:sin_c}) becomes
\begin{equation*}
    f_{k+2} = f_{k+1}+f_k \, .
\end{equation*}
Substituting Equation (\ref{eq:fk}) into this expression and solving for $c$, we find
\begin{equation*}
    c^2 - c - 1 = 0
\end{equation*}
with solution
\begin{equation*}
    c = \dfrac{1 + \sqrt{5}}{2} = \phi \, ,
\end{equation*}
the golden ratio. The same solution holds for all Equations (\ref{eq:sin}) with appropriate selection of $\{f_S,f_D,f_T\}$ from $\{f_k,f_{k+1},f_{k+2}\}$. We conclude that, for a system of damped coupled oscillators with oscillator and gain frequencies spaced by the multiplicative factor $c$, cross-frequency coupling between three neighboring rhythms requires $c = \phi$, the golden ratio. In other words, we propose that frequencies organized according to the golden ratio are particularly suited to support these cross-frequency interactions.
\\

\begin{figure}[h]
\centering
\includegraphics[width=14.5cm]{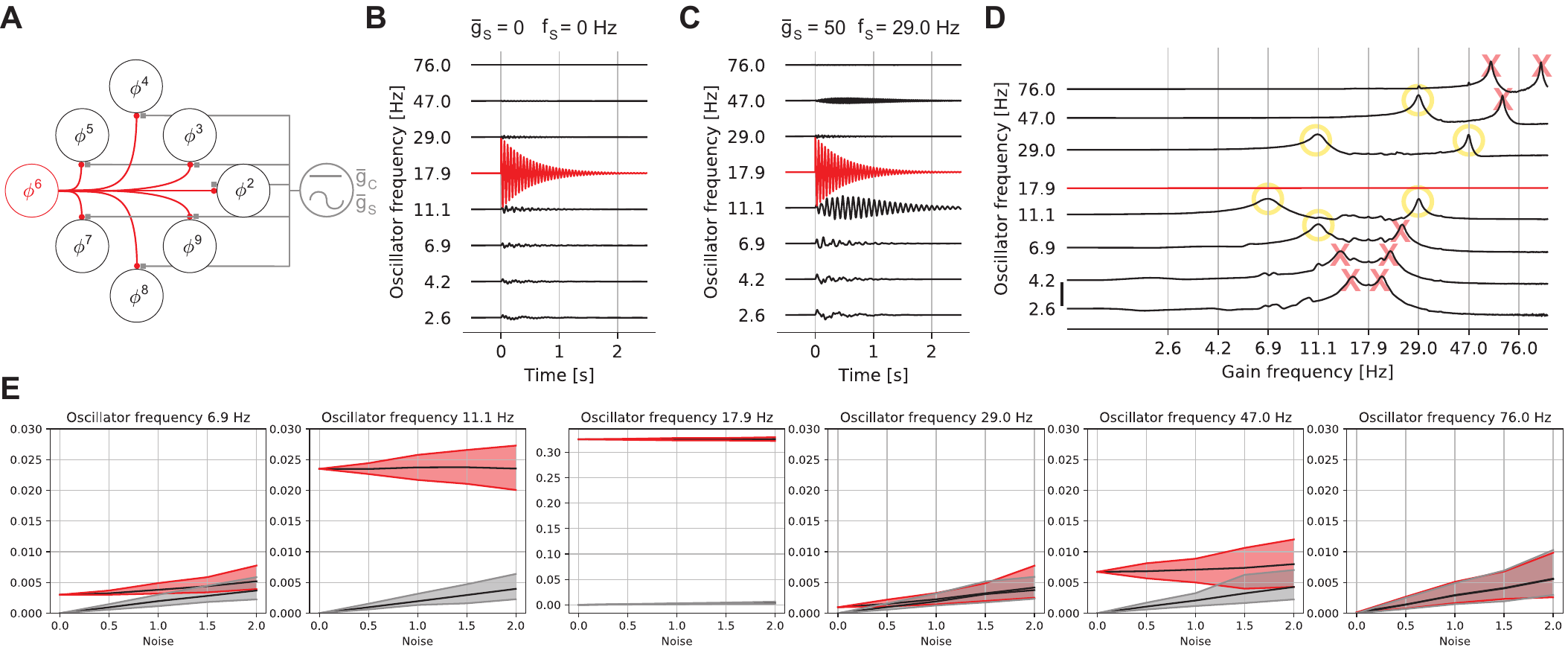}
\caption{\it{\textbf{Rhythms organized by the golden ratio support selective cross-frequency coupling.} \textbf{(A)} We perturb one oscillator (natural frequency 17.9 Hz, red), with connectivity to all other oscillators; $\phi$ is the golden ratio. \textbf{(B)} With only constant gain modulation, the perturbation ($t=0$, red) has little impact on other nodes. \textbf{(C)} With sinusoidal gain modulation at 29 Hz, two oscillators (natural frequencies 11.1 Hz and 47.0 Hz) selectively respond to the perturbation. \textbf{(D)} Average response amplitude (logarithm base 10) from $t=0$ to $t=1.5$ s at each oscillator versus gain frequency $f_S$. Different choices of gain frequency support selective coupling between the perturbed oscillator (natural frequency 17.9 Hz) and other oscillators. Peaks in response amplitude occur at golden rhythms (yellow circles, vertical lines) or other frequencies (red X’s). Minimum response set to 0 for each curve, and vertical scale bar indicates 1. \textbf{(E)} Average amplitude (black curve) and range ($2.5\%$ to $97.5\%$ from 100 simulations, shaded region) of evoked responses versus noise level. The oscillator with frequency $17.9$ Hz is directly perturbed, and sinusoidal gain modulation occurs with $f_S \approx 29.0$ Hz. Oscillators at golden rhythms exhibit different behavior with perturbation (red) versus without perturbation (gray). Code to simulate this network and create this figure is available \href{https://github.com/Mark-Kramer/Golden-Framework/blob/main/Figure-2.ipynb}{here}}.} \label{fig:8nodes_phi}
\end{figure}

\noindent To illustrate this result, we consider a network of 8 damped, coupled oscillators each with a different natural frequency determined by the golden ratio ($f_k=\phi^k$, where $\phi= \frac{(1+\sqrt{5})}{2}$; Figure \ref{fig:8nodes_phi}); we label these rhythms – scaled by factors of the golden ratio – as {\it golden rhythms}. Starting all nodes in a resting state, we perturb one oscillator ($f_D = \phi^6\approx17.9$~Hz) to produce a transient oscillation at that node. With only a constant gain $(\bar{g}_C=50, \bar{g}_S=0)$, the impact of the perturbation on the other oscillators is small (Figure \ref{fig:8nodes_phi}B); because $f_T \neq f_D$ for any oscillator pair, the network impact of the perturbation is small, despite the constant coupling.
\\

\noindent Including the sinusoidal gain modulation ($\bar{g}_C=50, \bar{g}_S=50$) results in selective communication between the oscillators. For example, choosing $f_S= \phi^7 \approx 29.0$ Hz, we observe an evoked response at two oscillators (Figure \ref{fig:8nodes_phi}C): $f_T= \phi^8 \approx 47.0$ Hz (consistent with Equation (\ref{eq:sin_a})) and $f_T= \phi^5 \approx 11.1$ Hz (consistent with Equation (\ref{eq:sin_c})). We note that the frequency of evoked responses matches the natural frequency of each oscillator. We also note that no solution exists for Equation (\ref{eq:sin_b}) because $f_T>0$. Different choices of gain frequency $f_S$ result in different pairs of cross-frequency coupling between the driver ($f_D$) and response oscillators (Figure \ref{fig:8nodes_phi}D). Cross-frequency coupling occurs when Equations (\ref{eq:sin}) are satisfied with $f_D \approx17.9$~Hz. The coupling is selective; for example, choosing a gain modulation of $f_S=11.1$ Hz results in cross-frequency coupling between the driver ($f_D=17.9$ Hz) and faster ($29$ Hz) and slower ($6.9$ Hz) golden rhythms. In this case, sinusoidal gain frequencies $f_S$ exist that support cross-frequency coupling and occur at factors of the golden ratio: i.e., $f_S = \phi^k$ (Figure \ref{fig:8nodes_phi}D, circles). We note that evoked responses also occur when $f_S \neq \phi^k$ (Figure \ref{fig:8nodes_phi}D, X’s); in these cases, frequencies outside the original rhythm sequence $f_k = \phi^k$ must exist to support cross-frequency coupling.  We conclude that if brain rhythmic activity – both oscillator and gain frequencies – organizes according to the golden ratio, then cross-frequency coupling is possible between a subset of separate rhythmic communication channels.

\noindent We now consider the impact of noise on this cross-frequency communication. With sinusoidal gain modulation ($\bar{g}_C=50, \bar{g}_S=50$, and $f_S= \phi^7 \approx 29.0$ Hz) and including noise in the oscillator dynamics (see {\it Methods}), we show the results for two cases: with perturbation and without perturbation to one oscillator ($f_D \approx 17.9$ Hz, as above). Without perturbation (gray in Figure \ref{fig:8nodes_phi}E), we find no evidence of an evoked response at any node, as expected; the amplitude remains small at all nodes, with a small gradual increase as the noise increases. With the perturbation (red in Figure \ref{fig:8nodes_phi}E), we find an evoked response at the perturbed oscillator ($f_D \approx 17.9$ Hz) and two other oscillators: $f_T \approx 47.0$ Hz (consistent with Equation (\ref{eq:sin_a})) and $f_T \approx 11.1$ Hz (consistent with Equation (\ref{eq:sin_c})). As the noise increases, so does the variability in the evoked response. For the lower frequency $f_T \approx 11.1$ Hz oscillator, the evoked response remains evident as the noise increases; in Figure \ref{fig:8nodes_phi}E, the perturbed (red) and unperturbed (gray) responses remain separate. For the higher frequency $f_T \approx 47.0$ Hz oscillator, the evoked response becomes more difficult to distinguish from the unperturbed case as the noise increases; in Figure \ref{fig:8nodes_phi}E, the perturbed (red) and unperturbed (gray) responses begin to overlap with increasing noise. We note that the amplitude of evoked responses decreases with frequency. Therefore, the same amount of noise impacts the higher frequency ($f_T \approx 47.0$ Hz) oscillator more than the lower frequency ($f_T \approx 11.1$ Hz) oscillator, making an evoked response more difficult to distinguish from background noise in the higher frequency case. We also note that oscillators not satisfying Equation (\ref{eq:sin}) (i.e., $f_T \approx \{6.9, 29.0, 76.0\}$ Hz when $f_D \approx 17.9$ Hz and $f_S \approx 29.0$ Hz) exhibit little evidence of an evoked response at any noise level.
\\

\begin{figure}[t]
\centering
\includegraphics[width=14.5cm]{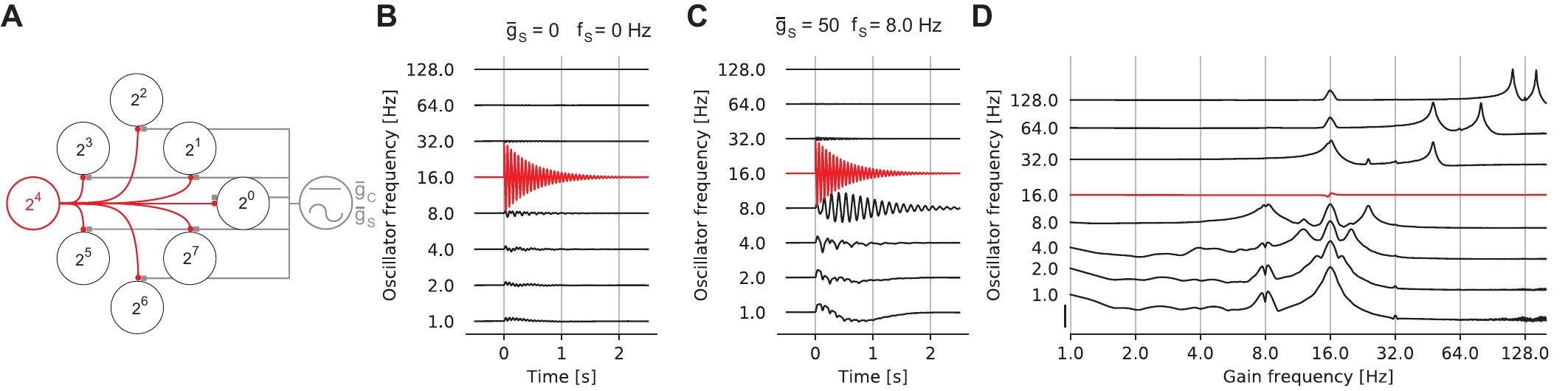}
\caption{\it{\textbf{An integer scaling between oscillators limits cross-frequency interactions.} \textbf{(A)} In a network of oscillators with frequencies organized by a factor of 2, we perturb one oscillator (natural frequency 16 Hz, red). \textbf{(B)} With constant gain, the impact of the perturbation is limited. \textbf{(C)} With sinusoidal modulation at $f_S=8$ Hz, a response appears at another oscillator (natural frequency 8 Hz). \textbf{(D)} The average response amplitude at each oscillator versus gain frequency $f_S$. Many oscillators respond when the gain frequency is 8 Hz, and responses tend not to occur at the oscillator frequencies; see Figure \ref{fig:8nodes_phi}D for plot details. Code to simulate this network and create this figure is available \href{https://github.com/Mark-Kramer/Golden-Framework/blob/main/Figure-3.ipynb}{here}}.} \label{fig:8nodes_2}
\end{figure}

\vspace{-0.075in}
\noindent To illustrate the utility of the golden ratio, we consider an alternative network of oscillators with frequencies organized by a factor of 2 (Figure \ref{fig:8nodes_2}A); such integer relationships have been proposed as important to neural communication \cite{palva_phase_2005,marin_garcia_genuine_2013,klimesch_algorithm_2013}. As expected, with only constant gain ($\bar{g}_C = 50$) a perturbation to one node ($f_D=16$ Hz) does not impact the rest of the network (Figure \ref{fig:8nodes_2}B). Including sinusoidal gain with frequency $f_S$ can produce cross-frequency coupling. For example, choosing $f_S=8$ Hz results in cross-frequency coupling between the $f_D=16$ Hz and $f_T=8$ Hz rhythms (Figure \ref{fig:8nodes_2}C). Similarly, choosing $f_S=16$ Hz results in cross-frequency coupling between the $f_D=16$ Hz and $f_T=32$ Hz rhythms; however, this choice of $f_S$ also results in strong cross-frequency coupling between $f_D=16$ Hz and lower frequency rhythms ($f_T=8,4,2,1$ Hz; Figure \ref{fig:8nodes_2}D). Importantly, we note that cross-frequency coupling typically occurs at sinusoidal gain frequencies that differ from the set of oscillator frequencies at $2^k$ Hz (vertical lines in Figure \ref{fig:8nodes_2}D); a new set of rhythms (and rhythm generators) must exist to support cross-frequency coupling in this network.
\\

\vspace{-0.075in}
\noindent To summarize, in a network of damped coupled oscillators (Equation \ref{eq:dsho}), sinusoidal gain modulation supports cross-frequency coupling (Equation \ref{eq:sin}). If oscillator and gain frequencies organize according to a multiplicative factor (Equation \ref{eq:fk}), then cross-frequency coupling between neighboring frequencies requires a multiplicative factor of $\phi$, the golden ratio (e.g., Figure \ref{fig:8nodes_phi}D). While oscillators organized with a different multiplicative factor can still produce cross-frequency coupling, the frequencies of effective gain modulation are not part of the original rhythmic sequence (e.g., Figure \ref{fig:8nodes_2}D), thus requiring the brain devote more resources to implementing a larger set of rhythms in support of cross-frequency interactions.

\subsection{Rhythms organized by the golden ratio support ensembles of cross-frequency coupling}

\begin{figure}[t]
\centering
\includegraphics[width=14.5cm]{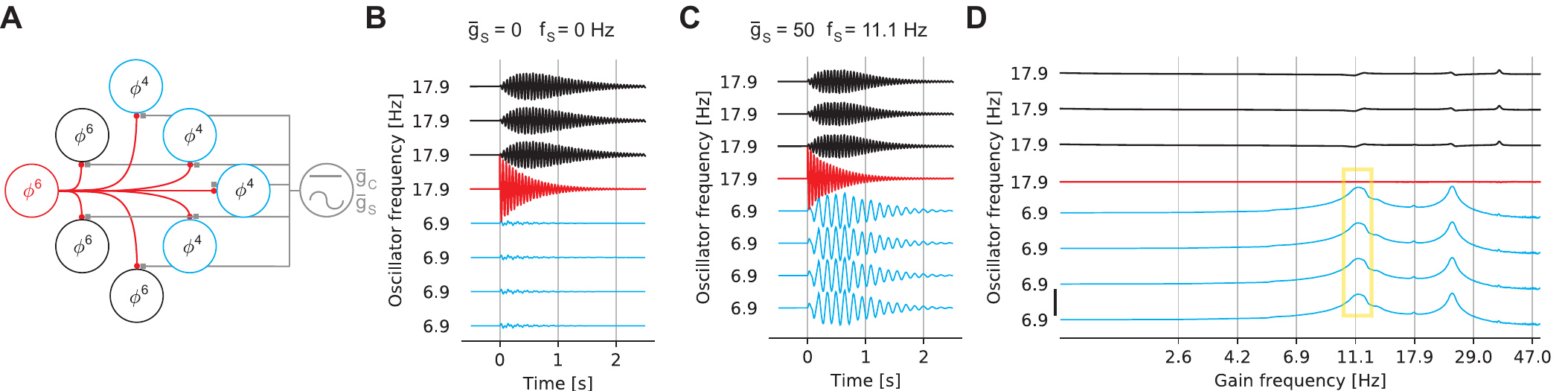}
\caption{\it{\textbf{Golden rhythms support coupling among an ensemble of nodes.} \textbf{(A)} The network consists of two ensembles with frequencies: $\phi^4$ and $\phi^6$. \textbf{(B)} With constant gain, perturbing a node in one ensemble impacts (red) other nodes in the same ensemble (black). \textbf{(C)} With sinusoidal gain at frequency $f_S=11.1$ Hz, the same perturbation impacts both ensembles. \textbf{(D)} Average amplitude response versus gain frequency for all nodes in both ensembles. The response at the unperturbed ensemble (blue) increases when the gain frequency is a golden rhythm (yellow box); see Figure \ref{fig:8nodes_phi}D for additional plot details. Code to simulate this network and create this figure is available \href{https://github.com/Mark-Kramer/Golden-Framework/blob/main/Figure-4.ipynb}{here}}.} \label{fig:ensemble_phi}
\end{figure}

In the previous section, we considered a network of nodes oscillating at different natural frequencies. As an alternative example, we now consider a network with two ensembles of nodes oscillating at different frequencies. The two ensembles consist of nodes oscillating at frequencies $\phi^k$ or $\phi^{k+2}$, where $\phi$ is the golden ratio. With only constant gain, a perturbation to any node impacts only nodes of the same ensemble (i.e., with the same frequency). Including sinusoidal gain modulation with (intermediate) frequency $f_S= \phi^{k+1}$, a perturbation to any node impacts nodes in both ensembles. We illustrate this in the 8-node network with 4 nodes in each ensemble oscillating at natural frequencies $\phi^4 \approx 6.85$ Hz or $\phi^6 \approx 17.9$ Hz (Figure \ref{fig:ensemble_phi}A). With only constant gain ($\bar{g}_C=50, \bar{g}_S=0$), a perturbation to one $\phi^6 \approx 17.9$ Hz (driver) node impacts the amplitude of all other nodes in the same ensemble (Figure \ref{fig:ensemble_phi}B). Including sinusoidal gain modulation ($\bar{g}_C=50, \bar{g}_S=50$) with frequency $\phi^5 \approx 11.1$ Hz, the same perturbation now impacts all nodes in both ensembles (Figure \ref{fig:ensemble_phi}C). From Equation \ref{eq:sin} we determine that two sinusoidal gain frequencies support cross-frequency coupling between the driver ($f_D= \phi^6 \approx 17.9$ Hz) and target ($f_T=\phi^4 \approx 6.85$ Hz) ensembles,
\begin{subequations}
\begin{eqnarray*}
    f_S=f_T-f_D=6.85-17.9<0.00 \, , \\
    f_S=f_D-f_T=17.9-6.85=11.1 \, \mathrm{Hz} \, , \\
    f_S=f_D+f_T=17.9+6.85=24.6 \, \mathrm{Hz} \, .
\end{eqnarray*}
\end{subequations}
\noindent However, of these two frequencies, only the former ($f_S=11.1$ Hz) is also a golden rhythm (Figure \ref{fig:ensemble_phi}D, box). In this case, cross-frequency coupling occurs when ensemble and gain rhythms organize in a “golden triplet” $(f_T, f_S, f_D) =(\phi^k,\phi^{k+1},\phi^{k+2}) \approx (6.85, 11.1, 17.9)$ Hz, where $\phi^k+\phi^{k+1} = \phi^{k+2}$.

\begin{figure}[h]
\centering
\includegraphics[width=14.5cm]{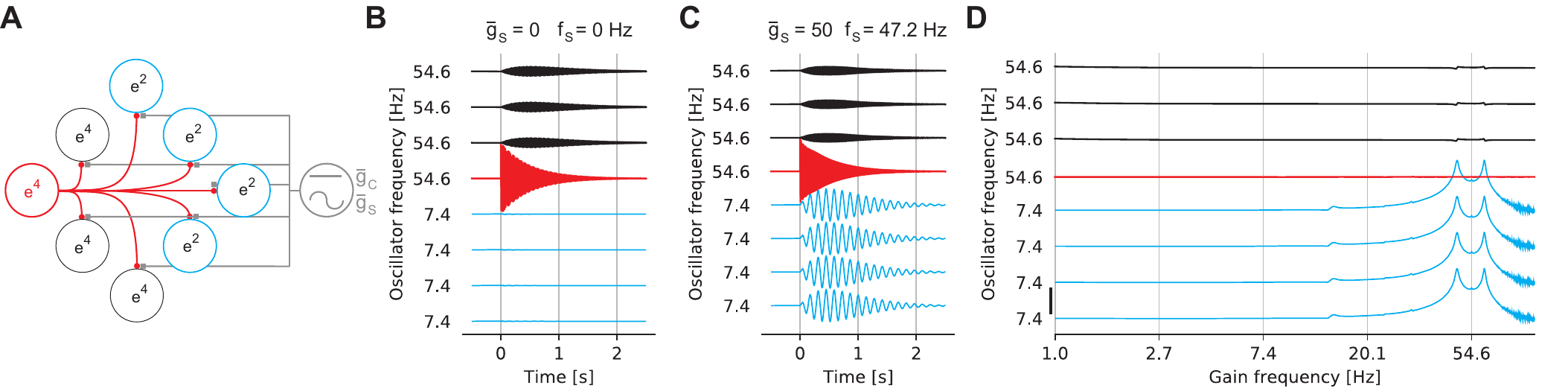}
\caption{\it{\textbf{Rhythms organized by Euler’s number do not support coupling between ensembles of nodes.} \textbf{(A)} The network consists of two ensembles with frequencies: $e^4$ and $e^2$. \textbf{(B)} With constant gain, perturbing a node in one ensemble impacts (red) other nodes in the same ensemble (black). \textbf{(C)} With sinusoidal gain at frequency $f_S=47.2$ Hz, the same perturbation impacts both ensembles. \textbf{(D)} Average amplitude response versus gain frequency for all nodes in both ensembles. The response at the unperturbed ensemble (blue) does not increase when the gain frequency is a factor of the Euler number (black vertical lines); see Figure \ref{fig:8nodes_phi}D for additional plot details. Code to simulate this network and create this figure is available \href{https://github.com/Mark-Kramer/Golden-Framework/blob/main/Figure-5.ipynb}{here}}.} \label{fig:ensemble_e}
\end{figure}

\noindent An alternative choice of irrational frequency ratio between the brain’s rhythms is Euler’s number ($e$) \cite{penttonen_natural_2003}. Repeating the simulation with two ensembles of frequency $e^k$ or $e^{k+2}$ results in cross-frequency coupling between ensembles only when $f_S=e^{k+2} \pm e^k$ (see Figure \ref{fig:ensemble_e} for an example with $k=2$). We therefore find similar results for the “Euler triplet” $(f_D, f_T,f_S)=(e^{k+2},e^k,e^{k+2} \pm e^k)$ or specifically for $k=2$, $(f_D, f_T, f_S)=(e^4,e^2,e^4 \pm e^2)$. However, this Euler triplet is not consistent with the ratio of $e$ observed {\it in vivo}, where three neighboring frequency bands appear at multiplicative factors of $e$ (e.g., $(f,ef,e^2f)$) and the two slower rhythms do not sum to equal the faster rhythm (e.g., $f+ef \neq e^2f)$. Only for three neighboring frequency bands related by the golden ratio $(f,\phi f, \phi^2f)$ do the frequencies of the slower rhythms sum to the faster rhythm (i.e., $f+\phi f=\phi^2f$).
\\

\subsection{Golden rhythms establish a hierarchy of cross-frequency interactions}
We now consider results derived for weakly coupled oscillators, which motivated the study of (strongly) coupled damped harmonic oscillators presented above. In \cite{izhikevich_weakly_1997,hoppensteadt_thalamo-cortical_1998}, Hoppensteadt and Izhikevich consider the general case of intrinsically oscillating neural populations with weak synaptic connections. When uncoupled, each neural population exhibits periodic activity (i.e., a stable limit cycle attractor) described by the phase of oscillation. We note that, in our study of coupled damped harmonic oscillators, we instead consider the amplitude of each oscillator. When Hoppensteadt and Izhikevich include weak synaptic connections between the neural populations, the phases of the neural populations interact only when a resonance relation exists between frequencies, i.e., 
\begin{equation*}
    \sum_{i}{k_i f_i}=0 ,
\end{equation*}
where $k_i$ is an integer and not all 0, and $f_i$ is the frequency of neural population $i$. The resonance order is then defined as the summed magnitudes of the integers $k_i$,
\begin{equation*}
    \mathrm{resonance\ order}= \sum_i ||{k_i}|| .
\end{equation*}
For the case of two neural populations, if
\begin{equation*}
    k_1 f_1+k_2 f_2 = 0
\end{equation*}
for integers $k_1$ and $k_2$, then
\begin{equation*}
    \frac{f_2}{f_1}=-\frac{k_1}{k_2} = \mathrm{rational} .
\end{equation*}
In other words, if the frequency ratio $f_2/f_1$ of the two neural populations is rational (i.e., the ratio of two integers), then the neural populations may interact, with the strength of interaction decreasing as either $k_1$ or $k_2$ increases (i.e., stronger interactions correspond to smaller resonance orders)\footnote{See Proposition 9.14 of \cite{izhikevich_weakly_1997}}. Alternatively, if this frequency ratio is irrational,
\begin{equation*}
    \frac{f_2}{f_1} = \mathrm{irrational} ,
\end{equation*}
then the two neural populations behave as if uncoupled.
\\

\noindent Consistent with the results presented here, Hoppensteadt and Izhikevich show that golden triplets possess the lowest resonance order, and therefore the strongest cross-frequency coupling \cite{izhikevich_weakly_1997,hoppensteadt_thalamo-cortical_1998,izhikevich_weakly_1999}. However, other resonances exist due to the recursive nature of rhythms organized by the golden ratio. To illustrate these relationships, we consider a set of golden rhythms $\{f^k\}$ – rhythms organized by the golden ratio so that,
\begin{equation} \label{eq:hierarchy1}
    f_{k-1}+f_k=f_{k+1} \, ,
\end{equation}
where $k$ is an integer. Because
\begin{equation*}
    f_{k-1}+f_k-f_{k+1}=0
\end{equation*}
the resonance order is $3$; this golden triplet supports strong cross-rhythm communication. Replacing $k$ with $k-1$ in Equation \ref{eq:hierarchy1}, we find
\begin{equation}\label{eq:hierarchy2}
    f_{k-2}+f_{k-1}=f_k \, .
\end{equation}
Then, replacing $f_k$ in Equation \ref{eq:hierarchy1} with the expression in Equation \ref{eq:hierarchy2}, we find
\begin{equation*}
    f_{k-1}+\big(f_{k-2}+f_{k-1}\big)=f_{k+1}
\end{equation*}
or
\begin{equation} \label{eq:heirarchy3}
    f_{k-2}+2f_{k-1}-f_{k+1}=0 \, ,
\end{equation}
which has resonance order $4$. Continuing this procedure to replace $f_{k-2}$ in the equation above, we find
\begin{equation} \label{eq:heirarchy4}
    {-f}_{k-3}+{3\ f}_{k-1}{-\ f}_{k+1}=0 \, ,
\end{equation}
which has resonance order $5$. In this way, golden rhythms support specific patterns of preferred coupling between rhythmic triplets, with the strongest coupling (lowest resonance order) between golden triplets.

\begin{figure}[h]
\centering
\includegraphics[width=5.5cm]{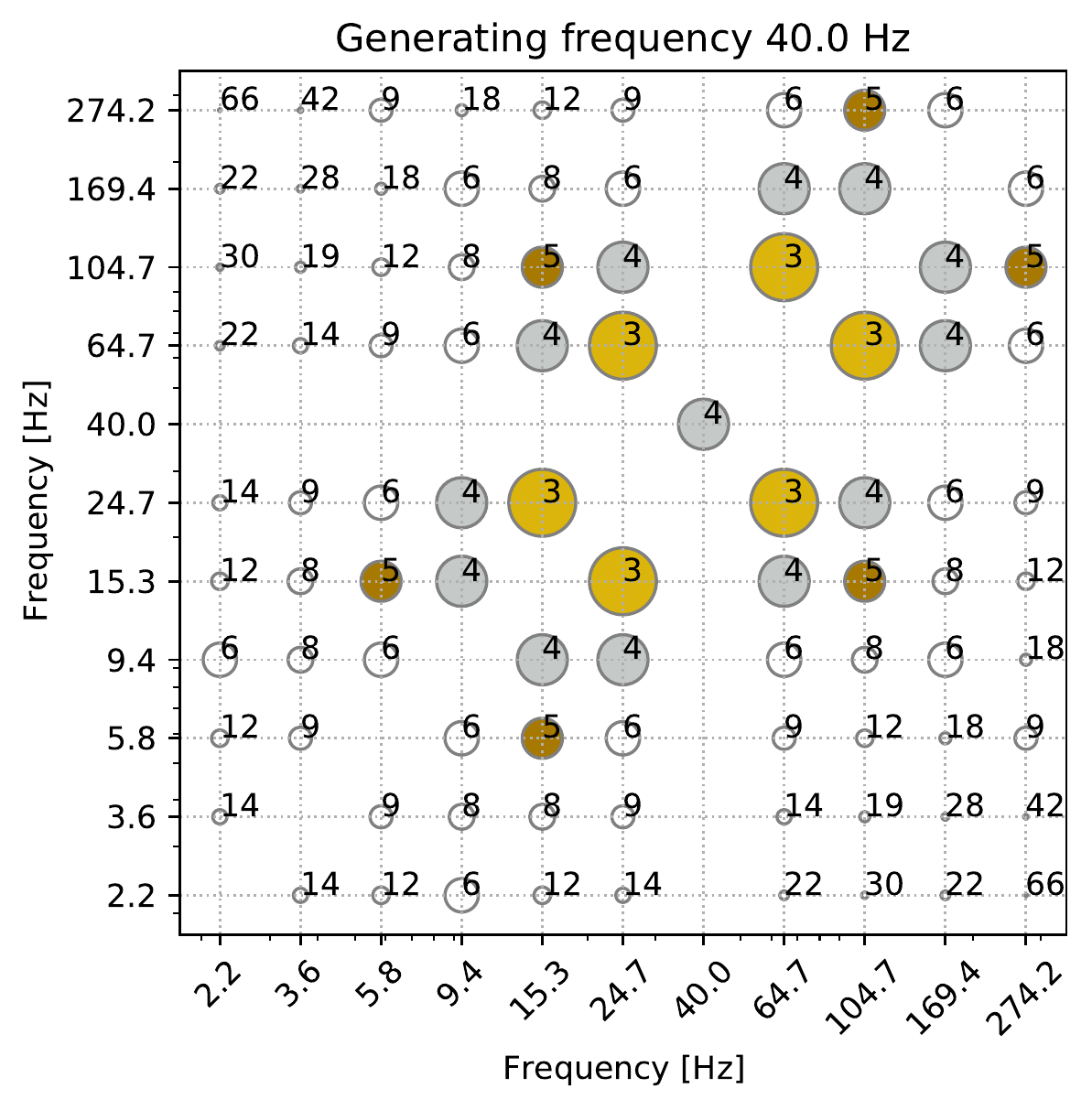}
\caption{\it{\textbf{Golden rhythms establish triplets with a discrete set of resonance orders.} The resonance order (numerical value, marker size) for triplets generated from 40 Hz. Lower resonance orders (3,4,5) indicated in color (gold, silver, bronze, respectively). Code to simulate this network and create this figure is available \href{https://github.com/Mark-Kramer/Golden-Framework/blob/main/Figure-6.ipynb}{here}}.} \label{fig:resonance_order}
\end{figure}

\noindent As a specific example, we fix $f_{k+1}=40$ Hz and list in Table \ref{tab:ex} the sequence of golden rhythms beginning with this generating frequency. We expect strong coupling between $(f_{k-1},f_k,f_{k+1}) = (15.3, 25, 40)$ Hz, a golden triplet, which has resonance order $3$. Using Equations \ref{eq:heirarchy3}, \ref{eq:heirarchy4}, and Table \ref{tab:ex}, we compute additional triplets with higher resonance orders: $(9.4, 15.3, 40)$ Hz with resonance order $4$, and $(5.8,15.3,40)$ Hz with resonance order $5$. Continuing this procedure organizes golden rhythms into triplets with different resonance orders (Figure \ref{fig:resonance_order}). Triplets with low resonance order appear near the target frequency of $f_{k+1}=40$ Hz (see gold, silver, and bronze circles in Figure \ref{fig:resonance_order}), and resonance orders tend to increase for frequencies further from $f_{k+1}=40$ Hz, with exceptions (e.g., $( f_{k-1},f_k, f_{k+1})=(2.2, 9.4, 40)$ Hz has resonance order $6$). We conclude that - based on theory developed for weakly coupled oscillators - golden rhythms support both separate communication channels and a hierarchy of cross-frequency interactions between rhythmic triplets with varying coupling strengths. While here we consider three interacting rhythms, we note that the theory also applies to four (or more) interacting rhythms. The implications of these results for networks of (strongly) coupled (damped) oscillators remains unclear.

\begin{table}[tbh]
\caption{{\it {\bf Example sequence of golden rhythms.} Beginning with $f_{k+1}=40$ Hz we compute the sequence of golden rhythms by multiplying or dividing by the golden ratio.}}
\label{tab:ex}
\centering
\begin{threeparttable}
\begin{tabular}{cccccc||c||cccc}
\headrow
\thead{$f_{k-5}$} & \thead{$f_{k-4}$} & \thead{$f_{k-3}$} & \thead{$f_{k-2}$} & \thead{$f_{k-1}$} & \thead{$f_{k}$} & \thead{$f_{k+1}$} & \thead{$f_{k+2}$} & \thead{$f_{k+3}$} & \thead{$f_{k+4}$} & \thead{$f_{k+5}$} \\
2.2 & 3.6 & 5.8 & 9.4 & 15.3 & 24.7 & 40 & 64.7 & 104.7 & 169.4 & 274.2 \\
\hline  % Please only put a hline at the end of the table
\end{tabular}
\end{threeparttable}
\end{table}

\subsection{Four experimental hypotheses}
We propose that golden rhythms optimally support separate and integrated communication channels between oscillatory neural populations. We now describe four hypotheses deduced from this theory. First, if the organization of brain rhythms follows the golden ratio, then we expect a discrete sequence of three frequency bands subdivides the existing gamma frequency band, broadly defined from 30-100 Hz \cite{buzsaki_neuronal_2004,fries_gamma_2007}, with peak frequencies separated by a factor of $\phi$. For example, using the sequence of golden rhythms with generating frequency 40 Hz (Table \ref{tab:ex}), we identify multiple distinct rhythms (at 40 Hz, 65 Hz, 105 Hz) corresponding to this gamma band. Consistent with this hypothesis, multiple distinct rhythms have been identified within the gamma band (e.g., \cite{lopes-dos-santos_parsing_2018,zhang_sub-second_2019,fernandez-ruiz_entorhinal-ca3_2017,edwards_high_2005,crone_functional_1998,vidal_visual_2006,colgin_frequency_2009,colgin_slow_2015,zhou_methodological_2019}). While different choices of generating frequency produce quantitatively different results, the qualitative result is the same: organized according to the golden ratio, multiple distinct rhythms exist within the gamma frequency range, each capable of supporting a separate communication channel.
\\

\begin{figure}[h]
\centering
\includegraphics[width=14.5cm]{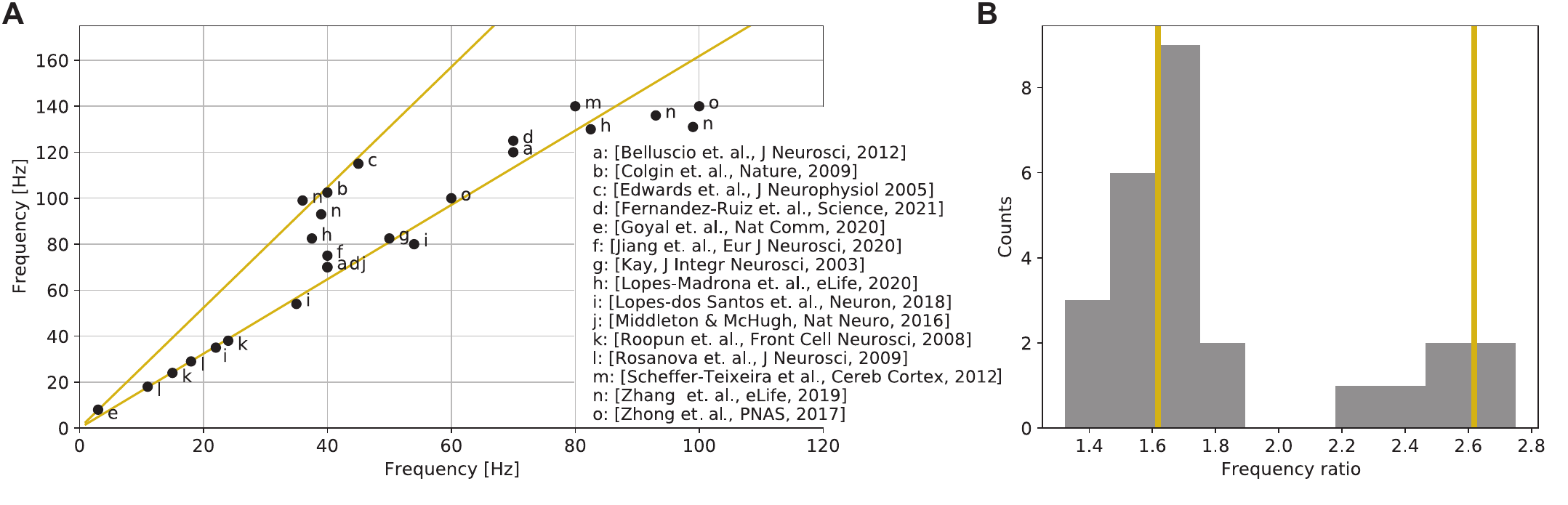}
\caption{\it{\textbf{Empirical observations of rhythms organized by the golden ratio in vivo.} \textbf{(A)} Pairs of frequencies reported in the literature; see legend. When only a frequency band is reported, we select the mean frequency of the band. \textbf{(B)} Histogram of the frequency ratio for each point in (A). Lines (golden) indicate frequency bands organized by $\phi \approx 1.6$ or $\phi^2 \approx 2.6$. Code to create this figure is available \href{https://github.com/Mark-Kramer/Golden-Framework/blob/main/Figure-7.ipynb}{here}}.} \label{fig:empirical}
\end{figure}

\noindent Second, if rhythms organize according to the golden ratio, then evidence for this relationship should exist {\it in vivo}. To that end, we consider examples of two or more frequency bands reported in the literature (predominately in rodent hippocampus; Figure \ref{fig:empirical}). These preliminary observations suggest that, in these cases, frequency bands separated by a factor of $\phi$ or $\phi^2$ commonly occur. 
\\

\noindent Third, if rhythms organize according to the golden ratio, then we propose that rhythmic triplets support cross-frequency communication. Nearly all existing research in cross-frequency coupling focuses on interactions between two rhythms (e.g., theta-gamma \cite{canolty_functional_2010,lisman_theta-gamma_2013,hyafil_neural_2015}), and many measures exist to assess and interpret bivariate coupling between rhythms \cite{hyafil_neural_2015,siebenhuhner_genuine_2020,tort_measuring_2010,aru_untangling_2015}. Yet, brain rhythms coordinate beyond pairwise interactions; trivariate interactions between three brain rhythms include coordination of beta, low gamma, and high gamma activity by theta phase \cite{belluscio_cross-frequency_2012,lopes-dos-santos_parsing_2018,zhang_sub-second_2019,fernandez-ruiz_entorhinal-ca3_2017,colgin_frequency_2009,jiang_distinct_2020}; coordination between ripples (140-200 Hz), sleep spindles (12-16 Hz), and slow oscillations (0.5-1.5 Hz) \cite{buzsaki_brain_2012}; and coordination between (top-down) beta, (bottom-up) gamma, and theta rhythms \cite{fries_rhythms_2015}. To assess trivariate coupling, an obvious initial choice is the bicoherence, which assesses the phase relationship between three rhythms: $f_1$, $f_2$, and $f_1+f_2$ \cite{barnett_bispectrum_1971,kramer_sharp_2008,shahbazi_avarvand_localizing_2018}. However, the bicoherence may be too restrictive (requiring a constant phase relationship between the three rhythms), and estimation of alternative interactions (e.g., between amplitudes and phases) will require application and development of alternative methods \cite{haufler_detection_2019}.
\\

\noindent Fourth, why brain rhythms occur at the specific frequency bands observed, and not different bands, remains unknown. To address this, we combine the golden ratio scaling proposed here with a fundamental timescale for life on Earth: the time required for Earth to complete one rotation (i.e., the sidereal period) of 23 hr, 56 min. Beginning from this fundamental frequency ($1/86160$ Hz), we compute higher frequency bands by repeated multiplication of the golden ratio (Table \ref{tab:earth}). Doing so, we identify frequencies consistent with the canonical frequency bands (i.e., delta, theta, alpha, beta, low gamma, middle gamma, high gamma, ripples, fast ripples; see last column of Table \ref{tab:earth}). We note that broad frequency ranges define the canonical frequency bands, for example the gamma band from (30, 100) Hz. Therefore, model predictions that identify rhythms within a band is not surprising. We propose instead that the relevant model prediction is the subdivision of the canonical frequency bands (e.g., the 10-30 Hz beta band into two sub-bands, the 30-100 Hz gamma band into three sub-bands), not the specific frequency values identified. We note that the lower frequencies may include "body oscillations", such as heart rate and breathing frequency \cite{tort_respiration-entrained_2018}, as proposed for a harmonic frequency relationship (factor of 2) in \cite{klimesch_algorithm_2013,klimesch_frequency_2018}. We hypothesize that if intelligent life were to evolve on a planet like Earth, in a star system like our own, with neural physiology like our own, then rhythmic bands would exist with center frequencies that depend on the planet’s circadian cycle. We acknowledge that this hypothesis, and the proposed association between neural rhythms and the sidereal period in Table \ref{tab:earth}, remain speculation, without robust supporting evidence.

\begin{table}[h]
\caption{{\it {\bf Golden rhythms, beginning with the sidereal period, align with the brain's rhythms.} The period ($T$) and frequency ($f$) of rhythms beginning with the sidereal period ($T=86160$ s) and multiplying the frequency by the golden ratio (number of multiplications indicated by the value in column {\bf Power}). Traditional frequency band labels (from \cite{buzsaki_neuronal_2004, buzsaki_scaling_2013, roopun_temporal_2008}) indicated in the last column.}}
\label{tab:earth}
\centering
\begin{threeparttable}
\begin{tabular}{ccc||ccc||ccc r}
\headrow
\thead{Power} & \thead{$T$ [s]} & \thead{$f$ [Hz]} & 
\thead{Power} & \thead{$T$ [s]} & \thead{$f$ [Hz]} &
\thead{Power} & \thead{$T$ [s]} & \thead{$f$ [Hz]} & Label \\
0&	86160&	1.16E-05&		12&	268&	0.004&	24&	0.83&	1.20&	Slow 1 \\
1&	53250&	1.88E-05&		13&	165&	0.006&	25&	0.51&	2&	Delta \\
2&	32910&	3.04E-05&		14&	102&	0.010&	26&	0.32&	3&	Delta \\
3&	20340&	4.92E-05&		15&	63.2&	0.016&	27&	0.20&	5&	Theta \\
4&	12571&	7.96E-05&		16&	39.0&	0.026&	28&	0.12&	8&	Alpha \\
5&	7769&	1.29E-04&		17&	24.1&	0.041&	29&	0.07&	13&	Beta1 \\
6&	4802&	2.08E-04&		18&	14.9&	0.067&	30&	0.05&	22&	Beta2  \\
7&	2968&	3.37E-04&		19&	9.2&	0.11&	31&	0.03&	35&	Low Gamma \\
8&	1834&	5.45E-04&		20&	5.7&	0.18&	32&	0.02&	57&	Mid Gamma \\
9&	1133&	8.82E-04&		21&	3.5&	0.28&	33&	0.01&	91&	High Gamma \\
10&	701&	0.001&		    22&	2.2&	0.46&	34&	0.01&	148&	Ripple \\
11&	433&	0.002&		    23&	1.3&	0.74&	35&	0.004&	239&	Fast Ripples \\
\hline  % Please only put a hline at the end of the table
\end{tabular}
\end{threeparttable}
\end{table}

\section{Discussion}
Why do brain rhythms organize into the small subset of discrete frequencies observed? Why does the alpha rhythm peak at 8-12 Hz and the (low) gamma rhythm peak at 35-55 Hz, across species \cite{buzsaki_scaling_2013}? Why does the brain not instead exhibit a continuum of rhythms, or a denser set of frequency bands, or different frequency bands? Here we provide a theoretical explanation for the organization of brain rhythms. Imposing a ratio of $\phi$ (the golden ratio) between the peaks of neighboring frequency bands, we constrain activity to a small subset of discrete brain rhythms, consistent with those observed {\it in vivo}. Organized in this way, brain rhythms optimally support the separation and integration of information in distinct rhythmic communication channels.
\\

\noindent The framework proposed here combines insights developed in existing works. Mathematical analysis of weakly coupled oscillators established the importance of resonance order for effective communication between neural populations oscillating at different frequencies \cite{izhikevich_weakly_1997,hoppensteadt_thalamo-cortical_1998,izhikevich_weakly_1999,nunez_brain_2010}. Experimental observations and computational models have established the importance of brain rhythms \cite{buzsaki_rhythms_2011,whittington_inhibition-based_2000,wang_neurophysiological_2010}, their interactions \cite{canolty_functional_2010,fries_rhythms_2015}, and their organization according to the golden ratio \cite{roopun_temporal_2008,kramer_rhythm_2008,roopun_period_2008}. Here, we combine these previous results with simulations and analysis of a network of damped, coupled oscillators in support of the proposed theory.
\\

\noindent The framework proposed here is consistent with existing theories for the role of brain rhythms. Like the communication-through-coherence (CTC) hypothesis \cite{fries_mechanism_2005,fries_rhythms_2015} and the frequency-division multiplexing hypothesis \cite{akam_oscillatory_2014,akam_efficient_2012}, in the framework proposed here neural populations communicate dynamically along anatomical connections via coordinated rhythms. Organization by the golden ratio complements these existing theories in two ways. First, by proposing which rhythms participate – namely, rhythms spaced by factors of the golden ratio. Second, by proposing the importance of three rhythms to establish cross-frequency interactions and proposing a hierarchical organization to these interactions.
\\

\noindent We considered a network of damped, coupled oscillators with sinusoidal gain modulation. In that network, cross-frequency coupling occurs when the gain frequency equals the sum or difference of the oscillator frequencies (Equation \ref{eq:sin}). This result holds without additional restrictions on the oscillator or gain frequencies. However, golden rhythms are unique in that oscillator and gain frequencies chosen from this set support cross-frequency coupling; no rhythms beyond this set are required. Alternative irrational scaling factors (e.g., Euler's number $e$) establish different sets of oscillator frequencies (e.g., $\ldots e^1, e^2, e^3, \ldots$) and separate communication channels, but require gain frequencies beyond this set to support cross-frequency coupling. In this alternative scenario, two distinct sets of rhythms exist: one reflecting local population activity, and another the cross-frequency coupling between populations. Rhythms organized by the golden ratio support a simpler framework: one set of frequencies (oscillator and gain) that reflect both local oscillations and their cross-frequency coupling. Golden rhythms are the smallest set of rhythms that support both separate communication channels and their cross-frequency interactions. Requiring fewer rhythms simplifies implementation, reducing the number of mechanisms required to produce these rhythms.
\\

\noindent These results are consistent with existing proposals that the golden ratio organizes brain rhythms and minimizes cross-frequency interference \cite{roopun_period_2008,weiss_golden_2003,pletzer_when_2010}. We extend these proposals by showing how triplets of golden rhythms facilitate cross-frequency coupling. An integer ratio of $2$ between frequency bands (with bandwidth determined by the golden ratio) provides an alternative organization to support cross-frequency coupling \cite{klimesch_algorithm_2013,klimesch_frequency_2018}. In this scenario, cross-frequency interactions have been proposed to occur via a shift in frequency. For example, two regions - with an irrational frequency ratio - remain decoupled until the center frequencies shift to establish $1:2$ phase coupling \cite{rodriguez-larios_mindfulness_2020,rodriguez-larios_thoughtless_2020}. We instead propose both regions maintain their original frequencies and couple when an appropriate third rhythm appears (e.g., a golden triplet). Our simulation results suggest more widespread coupling between populations oscillating at a $1:2$ frequency ratio compared to a golden ratio (Figure \ref{fig:8nodes_2}D). Interpreted another way, integer ratios between frequency bands may facilitate a "coupling superhighway"; a target region shifts frequency to enter the coupling superhighway and receive strong inputs from all upstream regions oscillating at integer multiples (or factors) of the target frequency. Rhythms organized by a golden ratio require coordination with a third input to establish cross-frequency coupling. Investigating these proposals requires analysis of larger networks with multiple rhythms, and perhaps multiple organizing frequency ratios.
\\

\noindent While we do not propose the specific mechanisms that support golden rhythms, proposals do exist. A biologically motivated sequence exists to create golden rhythms from the beta1 (15 Hz), beta2 (25 Hz), and gamma (40 Hz) bands. Through {\it in vitro} experiments and computational models, a process of period concatenation – in which the mechanisms producing the faster beta2 and gamma rhythms concatenate to create the slower beta1 rhythm – was proposed \cite{roopun_period_2008,roopun_temporal_2008,kramer_rhythm_2008}. Alternatively, golden rhythms may emerge when two input rhythms undergo a nonlinear transformation \cite{haufler_detection_2019,ahrens_spectral_2002}. The framework proposed here suggests that the emergent rhythms – appearing at the sum and difference of the two rhythms – may support local coordination of the input rhythms. 
\\

\noindent The simplicity of the proposed framework (compared to the complexity of brain dynamics) results in at least four limitations. First, brain rhythms appear as broad spectral bands, not sharply defined spectral peaks. Therefore, the meaning of a precise frequency ratio, or the practical difference between an irrational frequency ratio ($\phi$) and a rational frequency ratio (e.g., $1.6$) is unclear. Second, rhythm frequencies may vary systematically and continuously with respect to stimulus or behavioral parameters \cite{ray_differences_2010,kropff_frequency_2021}. Whether the brain maintains a constant frequency ratio between varying frequency rhythms, and what mechanisms could support this coordination, is unclear. Third, no evidence suggests the tuning of brain rhythms specifically to support separate communication channels. Instead, brain rhythms may occur at the frequencies observed due to the biological mechanisms available for coordination of neural activity (e.g., due to the decay time of inhibitory postsynaptic potentials that coordinate excitatory cell activity). Fourth, identification of rhythms in noisy brain signals remains a practical challenge, with numerous opportunities for confounds \cite{zhou_methodological_2019,cole_brain_2017,donoghue_methodological_2021}. Therefore, the best approach to compare this theory with data remains unclear.
\\

\noindent However, the simplicity of the golden framework is also an advantage. The framework consists of only one parameter (the golden ratio) compared to the many – typically poorly constrained – parameters of biologically detailed models of neural rhythms. In this way, the golden framework is broadly applicable and requires no specific biological mechanisms or rhythm frequencies; instead, only the relationship between frequencies is constrained.
\\

\noindent No theoretical framework exists to explain the discrete set of brain rhythms observed in nature. Here, we propose a candidate framework, simply stated: brain rhythms are spaced according to the golden ratio. This simple statement implies brain rhythms establish communication channels optimal for separate and integrated information flow. While the specific purpose of brain rhythms remains unknown, perhaps the brain evolved to these rhythms in support of efficient multiplexing on a limited anatomical network.

\section*{Acknowledgements}
The author would like to acknowledge Dr.\ Catherine Chu for writing assistance and tolerating many conversations about the golden ratio. 

\printendnotes

% Submissions are not required to reflect the precise reference formatting of the journal (use of italics, bold etc.), however it is important that all key elements of each reference are included.
\bibliography{golden-rhythms}

%\graphicalabstract{example-image-1x1}{Please check the journal's author guildines for whether a graphical abstract, key points, new findings, or other items are required for display in the Table of Contents.}

\appendix

\section{Relationship between a damped harmonic oscillator and autoregressive model of order two \label{Appendix1}}

Consider the damped harmonic oscillator driven by noise,
\begin{equation*}
    \ddot{x}_k + 2 \beta \dot{x}_k + \omega_k^2 x_k = \xi(t) \, ,
\end{equation*}
where $x$ is the position of the oscillator, $\beta$ is the damping constant, $\omega_0$ is the natural frequency, and $\xi(t)$ is a noise term evaluated at time $t$ (e.g., see Equation (5.28) of \cite{taylor_classical_2005}). Replacing each derivative with a discrete approximation we find,
\begin{equation*}
    \frac{x_t-2 x_{t-1}+x_{t-2}}{\Delta^2}+2 \beta \frac{x_t-x_{t-1}}{\Delta}+\omega_0^2\ x_t=\xi_t \, ,
\end{equation*}
where $x_t$ is the oscillator position at discrete time $t$, and $\Delta$ is the time between $t$ and $t+1$. Collecting terms at the same discrete time, we find,
\begin{equation*}
    \left(1+2\beta\Delta+\omega_0^2 \Delta^2\right) x_t -2 \left(1+\beta\Delta\right) x_{t-1}+x_{t-2}= \Delta^2 \xi_t \, , 
\end{equation*}
or
\begin{equation}\label{eq:AR}
    x_t= \alpha_1 x_{t-1}+\alpha_2 x_{t-2}+\epsilon_t,
\end{equation}
where
\begin{equation*}
\begin{array}{rl}
    \alpha_1 =& \dfrac{2 (1+\beta\Delta)}{1+2\beta\Delta+\omega_0^2 \Delta^2} \, , \\ \alpha_2 =& \dfrac{-1}{1+2\beta\Delta+\omega_0^2 \Delta^2} \, \\
    \epsilon_t =& \Delta^2 \xi_t \, .
\end{array}
\end{equation*}
Equation \ref{eq:AR} defines an autoregressive model of order 2 (i.e., an AR(2)).

\section{Resonance response for a damped, driven oscillator with sinusoidal gain \label{Appendix2}}

We begin with Equation \ref{eq:dsho},
\begin{equation} \label{eq:dshoB}
 \ddot{x}_k + 2 \beta \dot{x}_k + \omega_k^2 x_k = \left( \bar{g}_C + \bar{g}_S\cos{\omega_S t} \right) \sum_{j \neq k} x_j \ ,
\end{equation}
and simplify by replacing each $x_j$ with,
\begin{equation*}
    x_j \approx A_j\cos{ (\omega_j t)} \, ;
\end{equation*}
i.e., we assume each input oscillator $x_j$ oscillates with fixed amplitude ($A_j$) at its natural frequency ($\omega_j$). Then Equation \ref{eq:dshoB} becomes,
\begin{equation} \label{eq:dsho_sub_A}
    \ddot{x}_k + 2\beta \dot{x}_k + \omega_k^2 x_k = \bar{g}_C \sum_{j\neq k} A_j\cos{(\omega_j t)} + \bar{g}_S \cos{(\omega_S t)}\sum_{j\neq\ k}{A_j\cos{(\omega_j}t)} \, .
\end{equation}
Considering the first summation in Equation \ref{eq:dsho_sub_A} for the $j^{th}$ oscillator,
\begin{equation*}
    \ddot{x}_k + 2\beta \dot{x}_k + \omega_k^2 x_k = \bar{g}_C A_j\cos{(\omega_j t)}\, .
\end{equation*}
and applying the standard approach to solving a damped oscillator with sinusoidal driving force (e.g., see Chapter 5 of \cite{taylor_classical_2005}), we determine the amplitude $A_k$ of the driven oscillator,
\begin{equation}
    A_k^2 = \dfrac{\bar{g}_C^2 A_j^2}{(\omega_k^2-\omega_j^2)^2 + 4 \beta^2\omega_j^2} \, . 
\end{equation}
The amplitude of the driven oscillator is largest when $\omega_j=\omega_k$, i.e., when the frequency of the driving oscillator $\omega_j$ equals the natural frequency of the driven oscillator $\omega_k$.

\noindent We now consider the second summation in Equation \ref{eq:dsho_sub_A} for the $j^{th}$ oscillator,
\begin{equation*}
\begin{array}{rl}
    \ddot{x}_k + 2\beta \dot{x}_k + \omega_k^2 x_k =& \bar{g}_S A_j \cos{\left(\omega_j t\right)}\cos{\left(\omega_S t\right)} \, \\
    =& \dfrac{\bar{g}_S A_j}{2} \cos{\left( (\omega_j - \omega_S) t\right)}
    + \cos{\left( (\omega_j + \omega_S) t\right)} \, .
\end{array}
\end{equation*}
Any solution to this equation must also satisfy,
\begin{equation*}
\begin{array}{rl}
    \ddot{x}_k + 2\beta \dot{x}_k + \omega_k^2 x_k =&
    \dfrac{\bar{g}_S A_j}{2} \sin{\left( (\omega_j - \omega_S) t\right)}
    + \sin{\left( (\omega_j + \omega_S) t\right)} \, .
\end{array}
\end{equation*}
We define $z_k=x_k+i y_k$ and combine the two previous equations to find,
\begin{eqnarray*}
    \ddot{z}_k +2 \beta \dot{z}_k + \omega_k^2 z_k =&
     \dfrac{\bar{g}_S A_j}{2} e^{i\left(\omega_j-\omega_S\right)t}
    +\dfrac{\bar{g}_S A_j}{2} e^{i\left(\omega_j+\omega_S\right)t} \, .
\end{eqnarray*}
We now apply the standard approach to solving a damped oscillator with sinusoidal driving force (e.g., see Chapter 5 of \cite{taylor_classical_2005}) to determine the amplitude $B_k$ of the driven oscillator,
\begin{equation*}
    B_k^2= \dfrac{ \bar{g}_C^2 A_j^2 / 4}{\left(\omega_k^2-(\omega_j \pm \omega_S )^2 \right)^2 + 4 \beta^2 \left( \omega_j \pm \omega_S \right)^2} \, .
\end{equation*}
The amplitude $B_k$ is largest when,
\begin{equation*}
    \omega_k^2= \left(\omega_j \pm \omega_S \right)^2 \, ,
\end{equation*}
which is satisfied when,
\begin{equation}\label{eq:dsho_sub_soln1}
\begin{array}{ccc}
    \omega_j+\omega_S = \omega_k & \rightarrow & \omega_S = \omega_k-\omega_j \, , \\
    \omega_j-\omega_S = \omega_k & \rightarrow & \omega_S = \omega_j-\omega_k \, .
\end{array}
\end{equation}
Considering the equivalent expression,
\begin{eqnarray*}
    \omega_k^2 =& (-1)^2 \left(\omega_j \pm \omega_S \right)^2
               =& \left(-\omega_j \mp \omega_S \right)^2 \, ,
\end{eqnarray*}
we find an additional solution,
\begin{equation}\label{eq:dsho_sub_soln2}
\begin{array}{ccc}
    -\omega_j+\omega_S = \omega_k & \rightarrow & \omega_S= \omega_k+\omega_j \, .
\end{array}
\end{equation}
We conclude that the amplitude of the driven oscillator $B_k$ is largest when the frequency of the sinusoidal gain modulation $\omega_S$ equals the difference (Equation \ref{eq:dsho_sub_soln1}) or sum (Equation \ref{eq:dsho_sub_soln2}) of the natural frequencies of the driven $\omega_k$ and driving $\omega_j$ oscillators.

\end{document}